\documentclass[aps,prd,epsf,showpacs]{revtex4-1}
\pdfoutput=1
\usepackage[pdftex]{graphicx}
\usepackage{color}
\usepackage[dvipsnames]{xcolor}
\usepackage{amsmath}
\usepackage{amsfonts}
\usepackage{amssymb}
\usepackage{dsfont}
\usepackage[]{units}
\usepackage{braket}
\usepackage[utf8]{inputenc}
\usepackage[T1]{fontenc}
\usepackage{url}
\usepackage[english]{babel}
\usepackage{bm}% bold math
\usepackage{verbatim}
\usepackage{textcomp}
\usepackage{float}

\newcommand{\be}{\begin{equation}}
\newcommand{\ee}{\end{equation}}
\newcommand{\bea}{\begin{eqnarray}}
\newcommand{\eea}{\end{eqnarray}}

\newcommand{\bei}{\begin{itemize}}

\newcommand{\eei}{\end{itemize}}
\newcommand{\ben}{\begin{enumerate}}
\newcommand{\een}{\end{enumerate}}

\newcommand{\gtapprox}{\raisebox{-0.5ex}{$\,\stackrel{>}{\scriptstyle\sim}\,$}}
\newcommand{\ltapprox}{\raisebox{-0.5ex}{$\,\stackrel{<}{\scriptstyle\sim}\,$}}

\begin{document}

\title{Importance of meson-meson and of diquark-antidiquark creation operators for a $\bar{b} \bar{b} u d$~tetraquark}

\author{$^{(1)}$Pedro Bicudo}
\email{bicudo@tecnico.ulisboa.pt}

\author{$^{(2)}$Antje Peters}
\email{antje.peters@uni-muenster.de}

\author{$^{(3)}$Sebastian Velten}
\email{velten@itp.uni-frankfurt.de}

\author{$^{(3),(4)}$Marc Wagner}
\email{mwagner@itp.uni-frankfurt.de}

\affiliation{\vspace{0.1cm}$^{(1)}$CFTP, Dep.\ F\'{\i}sica, Instituto Superior T\'ecnico, Universidade de Lisboa, Av.\ Rovisco Pais, 1049-001 Lisboa, Portugal}

\affiliation{\vspace{0.1cm}$^{(2)}$Westf\"alische Wilhelms-Universit\"at M\"unster, Institut f\"ur Medizinische Psychologie und Systemneurowissenschaften, Von-Esmarch-Straße 52, D-48149 M\"unster, Germany}

\affiliation{\vspace{0.1cm}$^{(3)}$Johann Wolfgang Goethe-Universit\"at Frankfurt am Main, Institut f\"ur Theoretische Physik, Max-von-Laue-Stra{\ss}e 1, D-60438 Frankfurt am Main, Germany}

\affiliation{\vspace{0.1cm}$^{(4)}$Helmholtz Research Academy Hesse for FAIR, Campus Riedberg, Max-von-Laue-Stra{\ss}e 12, D-60438 Frankfurt am Main, Germany}

\begin{abstract}

In recent years, the existence of a hadronically stable $\bar{b} \bar{b} u d$ tetraquark with quantum numbers $I(J^P) = 0(1^+)$ was confirmed by first principles lattice QCD computations. 
In this work we use lattice QCD to compare two frequently discussed competing structures for this tetraquark by considering meson-meson as well as diquark-antidiquark creation operators.
We use the static-light approximation, where the two $\bar{b}$ quarks are assumed to be infinitely heavy with frozen positions, while the light $u$ and $d$ quarks are fully relativistic.
By minimizing effective energies and by solving generalized eigenvalue problems we determine the importance of the meson-meson and the diquark-antidiquark creation operators with respect to the ground state. 
It turns out, that the diquark-antidiquark structure dominates for $\bar{b} \bar{b}$ separations $r \ltapprox 0.25 \, \text{fm}$, whereas it becomes increasingly more irrelevant for larger separations, where the $I(J^P) = 0(1^+)$ tetraquark is mostly a meson-meson state. We also estimate the meson-meson to diquark-antidiquark ratio of this tetraquark and find around $60\% / 40\%$. 

\end{abstract}

% PACS, the Physics and Astronomy Classification Scheme.
% 12.38.Gc   Lattice QCD calculations (see also 11.15.Ha Lattice gauge theory).
% 13.75.Lb   Meson-meson interactions
% 14.40.Rt   Exotic mesons
% 14.65.Fy   Bottom quarks

\maketitle

%%%%%%%%%%%%%%%%%%%%%%%%%%%%%%%%%%%%%%%%%%%%%%%%%%%%%%%%%%%%%%%%%%%%%%%
%%
%%                                   SSS                             
%%                                  S   S                            
%%                                   S                               
%%                                    S                              
%%                                     S                             
%%                                  S   S                            
%%                                   SSS                             
%%
%%%%%%%%%%%%%%%%%%%%%%%%%%%%%%%%%%%%%%%%%%%%%%%%%%%%%%%%%%%%%%%%%%%%%%%

% \newpage

\section{\label{sec:intro}Introduction}

A long standing problem in particle physics is to understand exotic hadrons, i.e.\ hadrons which have a structure more complicated than a quark-antiquark pair or a triplet of quarks \cite{Jaffe:1976ig}. Such exotic hadrons turned out to be not only difficult to observe experimentally, but also technical to address in quark models \cite{Bicudo:2015bra}. In the last couple of years, several exotic hadrons, the majority pertaining to the class of tetraquarks with at least two heavy quarks, were clearly confirmed by the BELLE, BES-III and LHCb experimental collaborations. The observed exotic hadrons are resonances high in the spectrum. Studying them theoretically from first principles with lattice QCD requires the application and development of specific techniques from lattice hadron spectrosopy and scattering theory (see e.g.\ Refs.\ \cite{Prelovsek:2011nk,Morningstar:2016arm}).

There are two classes of doubly-heavy tetraquarks. Tetraquarks with one heavy quark and one heavy antiquark $\bar{Q} Q \bar{q} q$ including the $Z_c$ and $Z_b$ states are easier to detect experimentally. Their observation at Belle \cite{Belle:2011aa,Liu:2013dau,Chilikin:2014bkk}, Cleo-C \cite{Xiao:2013iha}, BESIII \cite{Ablikim:2013mio,Ablikim:2013emm,Ablikim:2013wzq,Ablikim:2013xfr,Ablikim:2014dxl} and LHCb \cite{Aaij:2014jqa} collaborations turned tetraquarks into a main highlight of particle physics in recent years. But since they are resonances with more than one decay channel, we study in this paper tetraquarks with two heavy antiquarks $\bar{Q} \bar{Q} q q$ (or equivalently two heavy quarks, i.e.\ $Q Q \bar{q} \bar{q}$), which are theoretically simpler, because they are either hadronically stable or can only decay into a pair of heavy-light mesons. Moreover, with the recent observation of hadronic systems with two heavy quarks \cite{Maciula:2016wci,Aaij:2017ueg} at LHCb we expect this second class of tetraquarks to be observed in the near future. Their discovery potential is discussed in Refs.\ \cite{Ali:2018ifm,Ali:2018xfq}.

These $\bar{Q} \bar{Q} q q$ tetraquarks are expected to form bound states, when the antiquarks are sufficiently heavy \cite{Ader:1981db,Ballot:1983iv,Heller:1986bt,Carlson:1987hh,Lipkin:1986dw,Brink:1998as,Gelman:2002wf,Vijande:2003ki,Janc:2004qn,Cohen:2006jg,Vijande:2007ix}. Recently this was confirmed with lattice QCD computations. One of the approaches uses the Born-Oppenheimer approximation \cite{Born:1927,Braaten:2014qka}, where the problem is split into two steps. The first step is to compute the potentials of two static antiquarks in the presence of two light quarks using state-of-the-art lattice QCD techniques (see e.g.\ Refs.\ \cite{Detmold:2007wk,Wagner:2010ad,Bali:2010xa,Wagner:2011ev,Brown:2012tm,Bicudo:2015kna}). Then, in the second step, the heavy quark dynamics is studied using a quantum mechanical Hamiltonian with the previously computed lattice QCD potentials. Using this approach, a $\bar{b} \bar{b} u d$ tetraquark bound state with quantum numbers $I(J^P) = 0(1^+)$ was predicted \cite{Bicudo:2012qt,Brown:2012tm,Bicudo:2015vta,Bicudo:2015kna,Bicudo:2016ooe}. Shortly afterwards, this was confirmed by several full lattice QCD computations using four quarks of finite mass \cite{Francis:2016hui,Francis:2018jyb,Junnarkar:2018twb,Leskovec:2019ioa,Hudspith:2020tdf}. 

Our present goal is to explore the structure of this $\bar{b} \bar{b} u d$ tetraquark with quantum numbers $I(J^P) = 0(1^+)$ using first principles lattice QCD computations, and to answer a hotly debated theoretical question
\cite{Bicudo:2004dx,Rai:2006wm,Chakrabarti:2008kc,Godfrey:2008nc,Lee:2008tz,Brambilla:2010cs,Faccini:2012pj,Esposito:2013fma,Guo:2013nja,Xie:2013uha,Esposito:2014hsa,Ignjatovic:2014wpj,Eichmann:2015cra,Brodsky:2015wza,Maiani:2016hxw,Esposito:2016itg,Rupp:2016jdk,Ali:2017jda,Sonnenschein:2018fph,Richard:2019cmi,Brambilla:2019esw,Maiani:2019lpu,Liu:2019mxw,Chen:2020aos,Yang:2020fou}: Is it a diquark-antidiquark system (denoted in the following as $Dd$) or rather a meson-meson system (denoted in the following as $BB$)?

The lattice QCD result for the static potential relevant for the $\bar{b} \bar{b} u d$ tetraquark with quantum numbers $I(J^P) = 0(1^+)$ can be parameterized by a screened Coulomb potential (see left plot of Figure~\ref{FIG900} and Refs.\ \cite{Detmold:2007wk,Wagner:2010ad,Wagner:2011ev,Brown:2012tm,Bicudo:2015kna}). This is consistent with the following.
(i) At $\bar{b} \bar{b}$ separations larger than the typical meson radius each of the two heavy antiquarks forms a bound state with one of the light quark. Thus we have a system composed of two separated $B$ mesons with interactions well-known to decay exponentially \cite{Gavela:1979zu} with Yukawa-like potentials \cite{Yukawa:1935xg}.
(ii) At small $\bar{b} \bar{b}$ separations the heavy antiquarks interact directly via gluon exchange and are immersed in a light quark cloud. Thus at small distances the potential is a Coulomb potential as expected from the asymptotic freedom of perturbative QCD (see e.g.\ Ref.\ \cite{Karbstein:2014bsa} and references therein). 

To compare quantitatively the importance of a $Dd$ structure and of a $BB$ structure for given $\bar{b} \bar{b}$ separation, we utilize a set of lattice QCD creation operators, both of $Dd$ type and of $BB$ type with static $\bar{b}$ quarks. A similar approach was previously used to explore systems of four quarks of finite mass. For example, studies of four-quark systems including a heavy $\bar{c} c$ pair were to some extent inconclusive. Neither clear evidence for the existence of a tetraquark was found, nor a signal improvement was observed for diquark-antidiquark operators \cite{Esposito:2013fma,Prelovsek:2014swa}. Also the tetraquark candidates $a_0(980)$ and the $D_{s0}^\ast(2317)$ were investigated in that way, employing sets of different creation operators, including quark-antiquark, meson-meson and/or diquark-antidiquark type \cite{Mohler:2013rwa,Lang:2014yfa,Bali:2017pdv,Alexandrou:2017itd,Alexandrou:2019tmk}. The focus of these studies was more to distinguish between a quark-antiquark and a four-quark structure, and since computations of tetraquark correlation functions, where all quarks have a finite mass, are extremely challenging \cite{Abdel-Rehim:2017dok}, no definite conclusion concerning meson-meson or diquark-antidiquark dominance was reached. On the other hand, studies of potentials and the corresponding gluon field distributions were performed with four static quarks, which are close to a system of four bottom quarks $\bar{b} \bar{b} b b$ \cite{Cardoso:2011fq,Cardoso:2012uka,Bicudo:2017usw}. In this case it was possible to clearly distinguish between a diquark-antidiquark structure and a meson-meson structure for the ground state depending on the geometric positions of the static sources.

This paper is structured as follows. In section~\ref{sec:lattice} we review important technical steps from our previous work \cite{Bicudo:2012qt} and discuss in detail the lattice QCD creation operators of $Dd$ type and of $BB$ type. We detail our lattice QCD setup in section~\ref{SEC100}. In Section~\ref{SEC499} we present our numerical results concerning the relative importance of a $Dd$ structure and of a $BB$ structure at given $\bar{b} \bar{b}$ separation. At the end of this section we use these results to crudely estimate the percentage of $Dd$ and of $BB$ in the $\bar{b} \bar{b} u d$ tetraquark with quantum numbers $I(J^P) = 0(1^+)$. We conclude in section~\ref{sec:conclusion}.

%%%%%%%%%%%%%%%%%%%%%%%%%%%%%%%%%%%%%%%%%%%%%%%%%%%%%%%%%%%%%%%%%%%%%%%
%%
%%                                   SSS                             
%%                                  S   S                            
%%                                   S                               
%%                                    S                              
%%                                     S                             
%%                                  S   S                            
%%                                   SSS                             
%%
%%%%%%%%%%%%%%%%%%%%%%%%%%%%%%%%%%%%%%%%%%%%%%%%%%%%%%%%%%%%%%%%%%%%%%%

% \newpage

\section{\label{sec:lattice}Potentials of two static antiquarks in the presence of two light quarks and corresponding creation operators}

In preceding papers \cite{Wagner:2010ad,Wagner:2011ev,Bicudo:2015vta,Bicudo:2015kna} we have computed potentials $V_{qq,j_z,\mathcal{P},\mathcal{P}_x}(r)$ of two static antiquarks $\bar{Q} \bar{Q}$ at separation $r$ in the presence of two light quarks $q q$ using lattice QCD. The computations have been carried out for many different sectors characterized by the following quantum numbers: light flavor $q q$ with $q \in \{ u , d, s, c \}$, total angular momentum of the light quarks and gluons $j_z$ with respect to the $\bar{Q} \bar{Q}$ separation axis, parity $\mathcal{P}$ and reflection along an axis perpendicular to the $\bar{Q} \bar{Q}$ separation axis $\mathcal{P}_x$. There are both attractive and repulsive sectors. Most promising with respect to the existence of $\bar{Q} \bar{Q} q q$ tetraquark bound states or resonances are attractive potentials with light quarks $q \in \{ u , d\}$, since they are rather wide and deep.

\begin{figure}[htb]
\includegraphics[width=0.49\columnwidth]{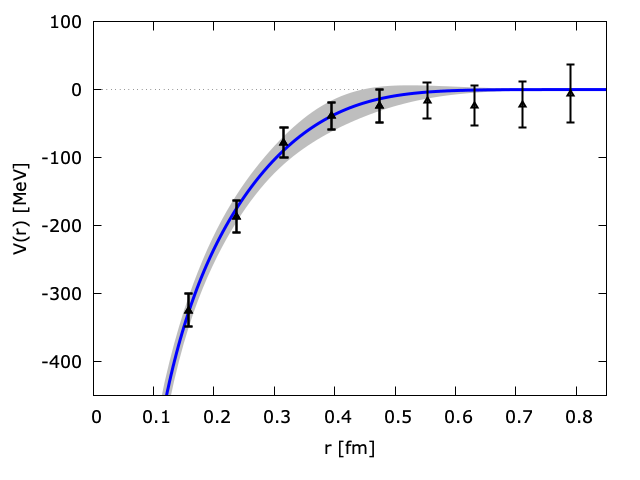}
\hfill
\includegraphics[width=0.49\columnwidth]{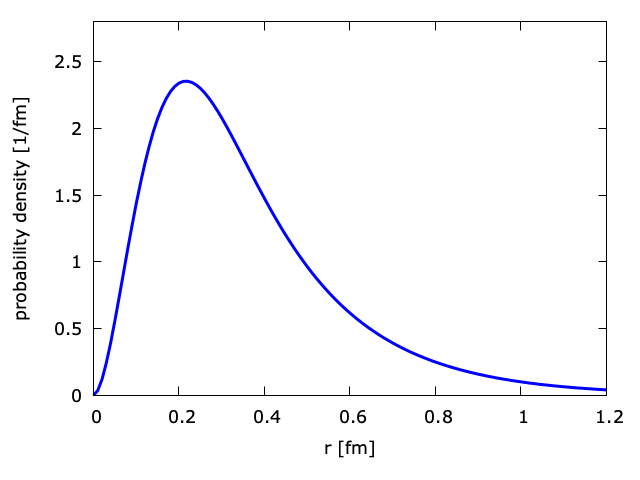}
\caption{\label{FIG900}\textbf{(left)}~Lattice QCD results for the potential $V(r) = V_{ud-du,0,-,+}(r)$ together with the parameterization $-(\alpha/r) e^{-(r/d)^p}$ with $\alpha = 0.293$, $d = 0.356 \, \textrm{fm}$ and $p = 2.74$. \textbf{(right)}~Probability density of the $\bar{b} \bar{b}$ separation $4 \pi |R(r)|^2$. (The results shown in the two plots are taken from Ref.\ \cite{Bicudo:2012qt}.)}
\end{figure}

Using the Born-Oppenheimer approximation, which amounts to solving the Schr\"odinger equation for the radial coordinate of the two heavy quarks $\bar{Q} \bar{Q} = \bar{b} \bar{b}$ with the computed potentials $V_{qq,j_z,\mathcal{P},\mathcal{P}_x}(r)$,
\begin{eqnarray}
\label{EQN649} \bigg(\frac{1}{m_b} \bigg(-\frac{d^2}{dr^2} + \frac{L (L+1)}{r^2}\bigg) + V_{qq,j_z,\mathcal{P},\mathcal{P}_x}(r) - 2 m_\text{sl}\bigg) R(r) = E R(r) ,
\end{eqnarray}
one can explore the existence of hadronically stable tetraquarks. $m_b$ is the $b$ quark mass, $L$ is the relative orbital angular momentum of $\bar{b} \bar{b}$ pair and $m_\text{sl}$ is the mass of the lightest static-light meson (computed within the same lattice QCD setup as $V_{qq,j_z,\mathcal{P},\mathcal{P}_x}(r)$; see e.g.\ Refs.\ \cite{Jansen:2008si,Michael:2010aa}). There is one particular potential $V(r) = V_{ud-du,0,-,+}(r)$ (shown in the left plot of Figure~\ref{FIG900}), which has quantum numbers $(I,j_z,\mathcal{P},\mathcal{P}_x) = (0,0,-,+)$, leading for $L = 0$ to a stable $\bar{b} \bar{b} u d$ tetraquark with quantum numbers $I(J^P) = 0(1^+)$ and binding energy $-E = 38(18) \, \textrm{MeV}$ \footnote{In our most recent and refined computation, which includes an extrapolation to physically light $u/d$ quark masses and a coupled channel Schr{\"o}dinger equation with a 2-component wave function, we find for the tetraquark mass $10 \, 545^{+38}_{-30} \, \textrm{MeV}$. This corresponds to binding energy $-E = 59^{+30}_{-38} \, \textrm{MeV}$ with respect to the $B B^\ast$ threshold (see Ref.\ \cite{Bicudo:2016ooe} for details).}. The probability density of the $\bar{b} \bar{b}$ separation $4 \pi |R(r)|^2$ (shown in the right plot of Figure~\ref{FIG900}) indicates that one typically finds separations in the range $0.1 \, \textrm{fm} \ltapprox r \ltapprox 0.6 \, \textrm{fm}$. None of the other potentials is sufficiently wide or deep to host a bound state \cite{Bicudo:2015vta}.

As discussed in the introduction, the main goal of this work is to investigate the structure of the predicted $\bar{b} \bar{b} u d$ tetraquark. In particular, we explore, whether the tetraquark is more similar to a meson-meson state $BB$ or to a diquark-antidiquark state $Dd$, two scenarios frequently discussed in the literature and at conferences also for other tetraquark candidates (see the discussion in section~\ref{sec:intro}). To this end, we refine the existing lattice QCD computation of the potential $V(r)$ by using two types of creation operators.

The first type of creation operators, which we used already in our previous computations \cite{Wagner:2010ad,Wagner:2011ev,Bicudo:2015vta,Bicudo:2015kna}, excites two $B$ mesons at separation $r$,
\begin{eqnarray}
\label{EQN_BB} \mathcal{O}_{BB,\Gamma} = 2 N_{BB}
(\mathcal{C} \Gamma)_{AB} (\mathcal{C} \tilde{\Gamma})_{CD} \Big(\bar{Q}_C^a(\mathbf{r}_1) \psi^{(f) a}_A(\mathbf{r}_1)\Big) \Big(\bar{Q}_D^b(\mathbf{r}_2) \psi^{(f') b}_B(\mathbf{r}_2)\Big)
\end{eqnarray}
with $r = |\mathbf{r}_2 - \mathbf{r}_1|$, color indices $a,b$, spin indices $A,B,C,D$ and $\psi^{(f)} \psi^{(f')} = u d - d u$. $N_{BB}$ is a normalization, which will be discussed later. There are two independent choices for the light spin matrix $\Gamma$ consistent with $(j_z,\mathcal{P},\mathcal{P}_x) = (0,-,+)$. $\Gamma = (1+\gamma_0) \gamma_5$ predominantly excites two negative parity ground state mesons $B^{(\ast)} B^{(\ast)}$, while $\Gamma = (1-\gamma_0) \gamma_5$ mostly generates two positive parity excited mesons $B_{0,1}^\ast B_{0,1}^\ast$, as one can see e.g.\ by applying a Fierz transformation to $\mathcal{O}_{BB,\Gamma}$ (see also Ref.\ \cite{Bicudo:2015kna}). Since static spins have no effect on energy levels, the heavy spin matrix is irrelevant and can be chosen arbitrarily, $\tilde{\Gamma} \in \{ (1-\gamma_0) \gamma_5 , (1-\gamma_0) \gamma_j \}$.

The second type of creation operators, which we use here for the first time, resembles a diquark-antidiquark pair with heavy quarks separated by $r$ and connected by a gluonic string,
\begin{eqnarray}
\label{EQN_Dd} \mathcal{O}_{Dd,\Gamma} = -N_{Dd}
\epsilon^{abc} \Big(\psi^{(f) b}_A(\mathbf{z}) (\mathcal{C} \Gamma)_{AB} \psi^{(f') c}_B(\mathbf{z})\Big)
\epsilon^{ade} \Big(\bar{Q}^f_C(\mathbf{r}_1) U^{fd}(\mathbf{r}_1;\mathbf{z}) (\mathcal{C} \tilde{\Gamma})_{CD} \bar{Q}^g_D(\mathbf{r}_2) U^{ge}(\mathbf{r}_2;\mathbf{z})\Big) .
\end{eqnarray}
Again $N_{Dd}$ is a normalization and the allowed light and heavy spin matrices are the same as for the operator $\mathcal{O}_{BB,\Gamma}$, i.e.\ $\Gamma \in \{ (1-\gamma_0) \gamma_5 , (1+\gamma_0) \gamma_5 \}$ and $\tilde{\Gamma} \in \{ (1-\gamma_0) \gamma_5 , (1-\gamma_0) \gamma_j \}$. Since the heavy spins are not part of the Hamiltonian, it is important to use the same $\tilde{\Gamma}$ for the operators $\mathcal{O}_{BB,\Gamma}$ and $\mathcal{O}_{Dd,\Gamma}$, whenever they are part of the same correlation matrix. For definiteness, we choose $\tilde{\Gamma} = (1-\gamma_0) \gamma_3$. $\mathbf{r}_1$ and $\mathbf{r}_2$ are always separated along one of the lattice axes and $\mathbf{z} = (\mathbf{r}_1 + \mathbf{r}_2) / 2$. For odd $r / a$ ($a$ denotes the lattice spacing), $\mathbf{z}$ does not coincide with one of the lattice sites. In these cases we take the average of the operator (\ref{EQN_Dd}) with $\mathbf{z} = (\mathbf{r}_1 + \mathbf{r}_2) / 2 + a \hat{\mathbf{r}} / 2$ and with $\mathbf{z} = (\mathbf{r}_1 + \mathbf{r}_2) / 2 - a \hat{\mathbf{r}} / 2$, where $\hat{\mathbf{r}} = (\mathbf{r}_2 - \mathbf{r}_1) / |\mathbf{r}_2 - \mathbf{r}_1|$.

We use smeared light quark and gluon fields, which implies that the operator $\psi_A^{(f) a}(\mathbf{r})$ does not generate a point-like excitation at $\mathbf{r}$, but rather a cloud-like excitation of spherical shape with diameter $\approx 0.5 \, \textrm{fm}$. Similarly, the gauge links connecting the two static antiquarks in the operator $\mathcal{O}_{Dd,\Gamma}$ generate a flux tube with a certain thickness. For details see section~\ref{SEC100}, where our lattice setup is discussed.

% see Eq. (27) in 0810.1843
% > sqrt((2 * 50 * 0.5) / (1 + 6*0.5)) * 2 * 0.079
% [1] 0.5586144
% > sqrt((2 * 50 * 0.5) / (1 + 6*0.5)) * 2 * 0.063
% [1] 0.4454773

Note that the creation operators $\mathcal{O}_{BB,\Gamma}$ and $\mathcal{O}_{Dd,\Gamma}$ do not generate orthogonal states, when applied to the vacuum $| \Omega \rangle$. For $r = 0$, the operators are even identical, when properly normalized, i.e.\ $\mathcal{O}_{BB,\Gamma} = \mathcal{O}_{Dd,\Gamma}$, if $N_{BB} = N_{Dd}$. This can easily be shown by using the identity $\epsilon^{abc} \epsilon^{ade} = \delta^{bd} \delta^{ce} - \delta^{be} \delta^{cd}$. For increasing $r$, however, they become more and more different, as we will show numerically in section~\ref{SEC486}. One obvious reason for that is that $\mathcal{O}_{Dd,\Gamma}$ creates a flux tube of length $r$, whereas $\mathcal{O}_{BB,\Gamma}$ does not. In section~\ref{SEC494}, section~\ref{SEC495} and section~\ref{SEC607} we will explore, whether the ground state of the $(I,j_z,\mathcal{P},\mathcal{P}_x) = (0,0,-,+)$ sector at a given separation $r$ is more similar to a $BB$ state $\mathcal{O}_{Dd,\Gamma} | \Omega \rangle$ or to a $DD$ state $\mathcal{O}_{Dd,\Gamma} | \Omega \rangle$. Since the energy of this ground state is the potential $V(r)$ used in the Born-Oppenheimer prediction of the $\bar{b} \bar{b} u d$ tetraquark with quantum numbers $I(J^P) = 0(1^+)$, our investigation will provide information on the structure of that tetraquark. We will discuss that in section~\ref{SEC600}.

%%%%%%%%%%%%%%%%%%%%%%%%%%%%%%%%%%%%%%%%%%%%%%%%%%%%%%%%%%%%%%%%%%%%%%%
%%
%%                                   SSS                             
%%                                  S   S                            
%%                                   S                               
%%                                    S                              
%%                                     S                             
%%                                  S   S                            
%%                                   SSS                             
%%
%%%%%%%%%%%%%%%%%%%%%%%%%%%%%%%%%%%%%%%%%%%%%%%%%%%%%%%%%%%%%%%%%%%%%%%

% \newpage

\section{\label{SEC100}Lattice QCD setup and techniques}

% **********

\subsection{Lattice actions}

The light quark action used in this work is the Wilson twisted mass action \cite{Frezzotti:2000nk,Frezzotti:2003ni},
\begin{eqnarray}
S_F[\chi,\bar{\chi},U] = \sum_x \bar{\chi}(x) \Big(D_W(m_0) + i \mu \gamma_5 \tau_3\Big) \chi(x) ,
\end{eqnarray}
where $D_W$ is the standard Wilson Dirac operator,
\begin{eqnarray}
D_W(m_0) = \frac{1}{2} \Big(\gamma_\mu (\nabla_\mu + \nabla^\ast_\mu) - \nabla^\ast_\mu \nabla_\mu\Big) + m_0 ,
\end{eqnarray}
with the forward and backward covariant derivatives $\nabla_{\mu}$ and $\nabla^\ast_{\mu}$. $\chi =
(\chi^{(u)} , \chi^{(d)})$ is the light quark doublet in the so-called twisted basis, which is related to quark fields in the usual ``physical basis'' via the twist rotation
\begin{eqnarray}
\label{EQN766} \psi = e^{i \gamma_5 \tau_3 \omega / 2} \chi ,
\end{eqnarray}
where $\omega$ is the twist angle. 

The gluon action used in this work is the tree-level Symanzik improved action \cite{Weisz:1982zw},
\begin{equation}
S_G[U] = \frac{\beta}{3} \sum_x \bigg(b_0 \sum_{\mu,\nu=1} \textrm{Re} \, \textrm{Tr}\Big(1 - P^{1 \times 1}_{\mu \nu}(x)\Big) + b_1 \sum_{\mu \neq \nu} \textrm{Re} \textrm{Tr}\Big(1 - P^{1 \times 2}_{\mu \nu}(x)\Big)\bigg) ,
\end{equation}
where $b_1 = -1/12$, $b_0=1-8b_1$, $\beta=6/g_0^2$, $g_0$ is the bare coupling and $P_{\mu \nu}^{1\times
1}$ and $P_{\mu \nu}^{1\times 2}$ are the plaquette and a $1 \times 2$ Wilson loop, respectively. 

To achieve automatic $\mathcal{O}(a)$ improvement, the hopping parameter $\kappa = (8+2 a m_0)^{-1}$ is tuned to its critical value, at which the PCAC quark mass vanishes \cite{Frezzotti:2000nk,Farchioni:2004ma,Farchioni:2004fs,Frezzotti:2005gi,Jansen:2005kk}. This corresponds to maximal twist, i.e.\ to tuning $\omega$ in
Eq.\ (\ref{EQN766}) to $\pi / 2$.

% **********

\subsection{Ensembles of gauge link configurations}

We have performed computations on two ensembles of gauge link configurations generated by the European Twisted Mass Collaboration (ETMC) \cite{Boucaud:2007uk,Boucaud:2008xu,Baron:2009wt}. These ensembles have different lattice spacings, $a \approx 0.079 \, \textrm{fm}$ and $a \approx 0.063 \, \textrm{fm}$ (set via the pion mass and pion decay constant and chiral perturbation theory \cite{Baron:2009wt}), which allows to study a finer resolution in the static antiquark-antiquark separation $r$ and to check for discretization errors. The pion masses $m_\textrm{PS}$ and spatial and temporal extents $L$ and $T$ are, however, very similar. The parameters and details of the ensembles are collected in Table~\ref{TAB540}.

\begin{table}[htb]
\begin{center}
\begin{tabular}{|c|c|c|c|c|c|c|c|}
\hline
 & & & & & & & \vspace{-0.40cm} \\
ensemble name & $\beta$ & $a$ in $\textrm{fm}$ & $(L/a)^3 \times T/a$ & $\kappa$ & $\mu$ & $m_\textrm{PS}$ in $\textrm{MeV}$ & \# configurations \\
 & & & & & & & \vspace{-0.40cm} \\
\hline
 & & & & & & & \vspace{-0.40cm} \\
\hline
 & & & & & & & \vspace{-0.40cm} \\
B40.24 & $3.90$ & $0.079(3)$ & $24^3 \times 48$ & $0.160856$ & $0.004$ & $340(13)$ & $108$ \\
 & & & & & & & \vspace{-0.40cm} \\
\hline
 & & & & & & & \vspace{-0.40cm} \\
C30.32 & $4.05$ & $0.063(2)$ & $32^3 \times 64$ & $0.157010$ & $0.003$ & $325(10)$ & $\phantom{0}98$ \\
\hline
\end{tabular}
\caption{\label{TAB540}Ensembles of gauge link configurations.}
\end{center}
\end{table}

% **********

\subsection{Smearing techniques}

As already mentioned in section~\ref{sec:lattice}, we have used several smearing techniques.

The spatial gauge links appearing in the $Dd$ creation operator (\ref{EQN_Dd}) are APE smeared \cite{Albanese:1987ds} with parameters $N_{\textrm{APE}} = 30$ and $\alpha_{\textrm{APE}} = 0.5$ (see Ref.\ \cite{Jansen:2008si}, Eq.\ (23)). The quark fields appearing in the $BB$ and $Dd$ creation operators (\ref{EQN_BB}) and (\ref{EQN_Dd}) are Gaussian smeared \cite{Gusken:1989qx} with parameters $N_{\textrm{Gauss}} = 50$ and $\kappa_{\textrm{Gauss}}=0.5$ (see Ref.\ \cite{Jansen:2008si}, Eq.\ (25)) and APE-smeared spatial gauge links. The intention of both APE and Gaussian smearing is to increase the ground state overlaps generated by the creation operators.

In addition to that we also use HYP2-smeared gauge links in temporal direction \cite{Hasenfratz:2001hp,DellaMorte:2003mn,DellaMorte:2005nwx} with parameters \\ $\alpha_1 = \alpha_2 = 1.0$ and $\alpha_3 = 0.5$. This reduces the self energy of the static quarks and, thus, improves the signal-to-noise ratio of the computed correlation functions.

%%%%%%%%%%%%%%%%%%%%%%%%%%%%%%%%%%%%%%%%%%%%%%%%%%%%%%%%%%%%%%%%%%%%%%%
%%
%%                                   SSS                             
%%                                  S   S                            
%%                                   S                               
%%                                    S                              
%%                                     S                             
%%                                  S   S                            
%%                                   SSS                             
%%
%%%%%%%%%%%%%%%%%%%%%%%%%%%%%%%%%%%%%%%%%%%%%%%%%%%%%%%%%%%%%%%%%%%%%%%

% \newpage

\section{\label{SEC499}Numerical results}

We consider the following four creation operators,
\begin{eqnarray}
\label{EQN868} \mathcal{O}_j \quad , \quad j \in \Big\{ [BB,(1 + \gamma_0) \gamma_5] \ , \ [BB,\gamma_5] \ , \ [Dd,(1 + \gamma_0) \gamma_5] \ , \ [Dd,\gamma_5] \Big\} ,
\end{eqnarray}
and define the corresponding trial states as
\begin{eqnarray}
\label{EQN841} | \Phi_j \rangle = \mathcal{O}_j | \Omega \rangle .
\end{eqnarray}
$\mathcal{O}_{BB,(1 + \gamma_0) \gamma_5}$ predominantly excites two negative parity ground state mesons. Thus, $| \Phi_{BB,(1 + \gamma_0) \gamma_5} \rangle$ is expected to have the largest overlap to the ground state of the $(I,j_z,\mathcal{P},\mathcal{P}_x) = (0,0,-,+)$ sector at large $r$. At small $r$, however, the diquark-antidiquark operators might be advantageous. We use $| \Phi_{Dd,\gamma_5} \rangle$, i.e.\ a light diquark with just $\gamma_5$, as typically discussed in the literature. Since $\mathcal{O}_{BB,\Gamma} \propto \mathcal{O}_{Dd,\Gamma}$ for $r = 0$ (see section~\ref{sec:lattice}), the diquark-antidiquark trial state $| \Phi_{Dd,(1 + \gamma_0) \gamma_5} \rangle$ might even be a better candidate for having a large ground state overlap. For completeness we also include $\mathcal{O}_{BB,\gamma_5}$, the mesonic molecule counterpart of $\mathcal{O}_{Dd,\gamma_5}$. $\mathcal{O}_{BB,\gamma_5}$ excites a linear combination of two negative parity ground state mesons and two significantly heavier positive parity excited mesons \cite{Bicudo:2015kna}.

With these operators we computed the $4 \times 4$ correlation matrix,
\begin{eqnarray}
\label{EQN755} C_{j k}(t) = \Big\langle \mathcal{O}^\dagger_j(t_2) \mathcal{O}_k(t_1) \Big\rangle = \langle \Omega | \mathcal{O}^\dagger_j(t_2) \mathcal{O}_k(t_1) | \Omega \rangle = \langle \Phi_j(t_2) | \Phi_k(t_1) \rangle
\end{eqnarray}
with $\langle \ldots \rangle$ denoting the path integral expectation value and $t/a = (t_2 - t_1)/a \geq 1$. For the second equality we assumed that in the spectral decomposition propagation over temporal separation $T-t$ is suppressed for all states except for the vacuum. To cross-check our computations, we checked the numerical results with respect to the symmetries $\gamma_5$ hermiticity, parity, time reversal, charge conjugation and cubic rotations around the axis of separation (for details see Ref.\ \cite{Bicudo:2015kna}). In a second step we averaged elements of the correlation matrices related by these symmetries, to reduce statistical errors.

% **********

\subsection{\label{SEC486}Squared overlaps of the normalized $BB$ and $Dd$ trial states}

In this subsection we study
\begin{eqnarray}
\alpha_{j k}(t) = \frac{|C_{j k}(t)|^2 }{C_{j j}(t) C_{k k}(t)} .
\end{eqnarray} 
For $t \rightarrow 0$ this quantity is the squared normalized overlap of trial state $| \Phi_j \rangle = \mathcal{O}_j | \Omega \rangle$ and trial state $| \Phi_k \rangle = \mathcal{O}_k | \Omega \rangle$, i.e.\
\begin{eqnarray}
\alpha^0_{j k} = \lim_{t \rightarrow 0} \alpha_{j k}(t) = \frac{|\langle \Phi_j | \Phi_k \rangle|^2}{\langle \Phi_j | \Phi_j \rangle \langle \Phi_k | \Phi_k \rangle} .
\end{eqnarray}
Clearly, $0 \leq \alpha^0_{j k} \leq 1$. Note that for an arbitrary state $| \Psi \rangle$ and an orthonormal basis $| k \rangle$, $k = 1,2,3,\ldots$
\begin{eqnarray}
\sum_{k = 1}^\infty \frac{|\langle \Psi | k \rangle|^2}{\langle \Psi | \Psi \rangle \langle k | k \rangle} = 1 .
\end{eqnarray}
Thus for two trial states $| \Phi_j \rangle$ and $| \Phi_k \rangle$, $\alpha^0_{j k}$ can be interpreted as a measure of their orthogonality, where $\alpha^0_{j k} \approx 0$ indicates almost orthogonal and $\alpha^0_{j k} \approx 1$ almost parallel states.
For large $t$ all $\alpha_{j k}$ approach $1$, because the ground state dominates in that limit.

\begin{figure}[htb]
\includegraphics[width=0.49\columnwidth]{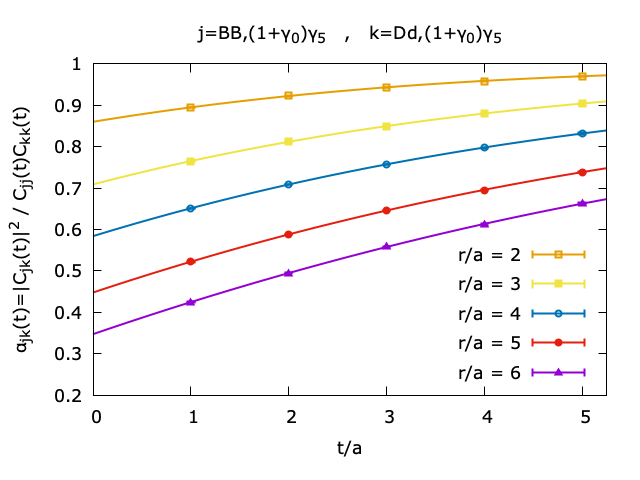}
\hfill
\includegraphics[width=0.49\columnwidth]{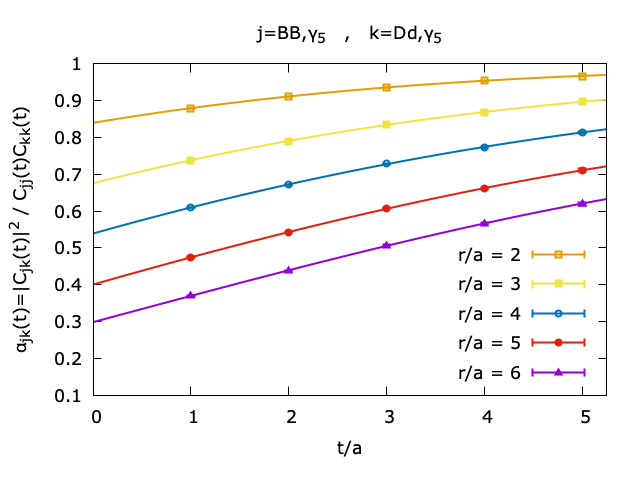}
\caption{\label{fig:orthog}$\alpha_{j k}$ as a function of $t$ for several fixed $r$ for ensemble B40.24. \textbf{(left)}~$j = BB,(1+\gamma_0) \gamma_5$, $k = Dd,(1+\gamma_0) \gamma_5$. \textbf{(right)}~$j = BB,\gamma_5$, $k = Dd,\gamma_5$. For $t \rightarrow 0$, $\alpha_{j k}$ is the squared overlap of the corresponding normalized trial states. The curves, which are fits with degree-2 polynomials, represent crude extrapolations to $t = 0$.}
\end{figure}

In the left plot of Figure~\ref{fig:orthog} we show $\alpha_{j k}$ for $j = BB, (1+\gamma_0) \gamma_5$ and $k = Dd, (1+\gamma_0) \gamma_5$ as function of $t$ for several fixed $r$ \footnote{Static-static correlation functions computed with lattice QCD exhibit strong discretization errors for separations $r/a \ltapprox 2$, as is e.g.\ well known in the context of computations of Wilson loops and the ordinary static potential. Thus, throughout this work we only show and discuss results for $r/a \geq 2$.} for ensemble B40.24. The corresponding trial states become more similar for smaller $r$, as indicated by larger values of $\alpha_{j k}$ close to $1$. This is not surprising, because for $r = 0$ the two operators $\mathcal{O}_{BB,\Gamma}$ and $\mathcal{O}_{Dd,\Gamma}$ are identical, if normalized according to $N_{BB} = N_{Dd}$, as discussed in section~\ref{sec:lattice}. For larger heavy quark separations $r$, the two trial states clearly differ. $\mathcal{O}_{BB,(1+\gamma_0) \gamma_5}$ generates a pair of spatially separated $B$ mesons, which interact only by weak residual hadronic forces. In contrast to that, the antiquarks $\bar{Q} \bar{Q}$ forming the heavy diquark in the operator $\mathcal{O}_{Dd,(1+\gamma_0) \gamma_5}$ are connected by a gluonic flux tube of length $r$. The data points for $1 \leq t/a \leq 5$ can be fitted consistently with degree-2 polynomials, which are also shown in Figure~\ref{fig:orthog}. These fits represent crude extrapolations of $\alpha_{j k}$ to $t = 0$, which are the squared overlaps of the corresponding normalized $BB$ and $Dd$ trial states.

In the right plot of Figure~\ref{fig:orthog} we show the corresponding results for $j = BB, \gamma_5$ and $k = Dd, \gamma_5$. They are quite similar to those for $j = BB, (1+\gamma_0) \gamma_5$ and $k = Dd, (1+\gamma_0) \gamma_5$ and can be interpreted in the same way. The main point of these two plots is to demonstrate that $BB$ and $Dd$ trial states, even though not orthogonal, are not linearly dependent either. In the following subsections we explore, whether the ground state of the $(I,j_z,\mathcal{P},\mathcal{P}_x) = (0,0,-,+)$ sector for given $r$ is more similar to a $BB$ trial state or to a $Dd$ trial state.

Analog plots of $\alpha_{j k}$ for ensemble C30.32 are very similar to those of Figure~\ref{fig:orthog} and, thus, not shown.

% **********

\subsection{\label{SEC494}Effective energies corresponding to diagonal elements of the correlation matrix at small and large temporal separations}

\begin{figure}[htb]
\includegraphics[width=0.49\columnwidth]{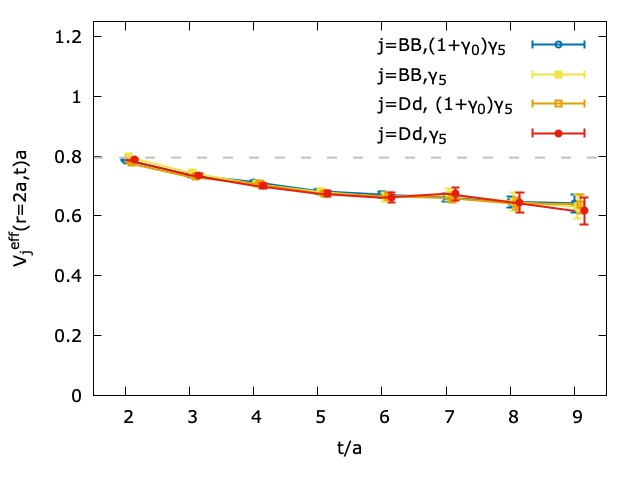}
\hfill
\includegraphics[width=0.49\columnwidth]{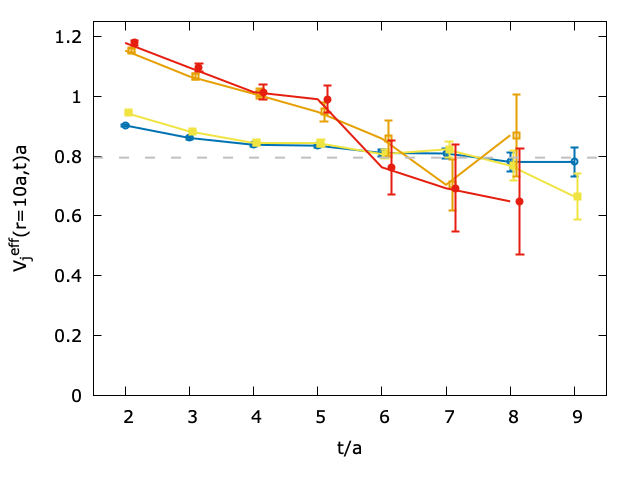}
\caption{\label{fig:Veff}Effective energies $V_j^\text{eff}$ corresponding to diagonal elements of the correlation matrix for fixed $r$ as functions of $t$ for ensemble B40.24. \textbf{(left)}~$r = 2 a \approx 0.16 \, \text{fm}$. \textbf{(right)}~$r = 10 a \approx 0.79 \, \text{fm}$. The dotted gray line in both plots represents the $BB$ threshold at $2 m_B$, where $m_B$ is the mass of the lightest static light meson taken from our previous work \cite{Michael:2010aa}.}
\end{figure}

Now we consider effective energies corresponding to diagonal elements of the correlation matrix, i.e.\
\begin{eqnarray}
\label{EQN539} V_j^\text{eff}(r,t) = -\frac{1}{a} \log\bigg(\frac{C_{jj}(t)}{C_{jj}(t-a)}\bigg) \quad \text{(no sum over } j \text{)} .
\end{eqnarray}
As discussed in section~\ref{sec:lattice}, all four operators probe the $(I,j_z,\mathcal{P},\mathcal{P}_x) = (0,0,-,+)$ sector. Thus, for fixed $r$ and at sufficiently large $t$, all four $V_j^\text{eff}(r,t)$ should approach the same constant, which is the ground state energy $V(r)$. For separations $r \ltapprox 0.3 \, \textrm{fm}$, our numerical results confirm that expectation (see e.g.\ the left plot of Figure~\ref{fig:Veff}, where $V_j^\text{eff}(r,t)$ is shown for $r \approx 0.16 \, \text{fm}$ as a function of $t$). For larger $r$, however, the two $Dd$ operators seem to generate only little overlap to the ground state. The consequence is that the corresponding effective energies $V_{Dd,(1 + \gamma_0) \gamma_5}^\text{eff}$ and $V_{Dd,\gamma_5}^\text{eff}$ do not convincingly converge to the plateau at $V(r)$ in the $t$ region, where we carried out computations, and have rather large statistical errors for larger $t$ separations (see e.g.\ the right plot of Figure~\ref{fig:Veff}, where $V_j^\text{eff}(r,t)$ is shown for $r \approx 0.79 \, \text{fm}$ as a function of $t$). In contrast to that, the two $BB$ operators still lead to clear effective energy plateaus. Thus, if one is just interested to determine $V(r)$, it is sufficient to implement $BB$ operators. This is, what we did in previous work, e.g.\ in Ref.\ \cite{Bicudo:2012qt} (see also the left plot of Figure~\ref{FIG900}).

% m(S   ) = 0.397484 +/- 0.002183   for B40.24   from Michael:2010aa
% m(S   ) = 0.332838 +/- 0.004358   for C30.32   from Michael:2010aa

A first indicator concerning the structure of the ground state are effective energies at small temporal separations, i.e.\ $V_j^\text{eff}(r,t=2a)$. Since there is little suppression of excited states by the Euclidean time evolution, a small value of $V_j^\text{eff}(r,t=2a)$ close to the ground state energy $V(r)$ implies an operator, which predominantly excites the ground state. A larger value of $V_j^\text{eff}(r,t=2a)$, on the other hand, is a sign that the corresponding operator creates a trial state less similar to the ground state.

\begin{figure}[htb]
\includegraphics[width=0.49\columnwidth]{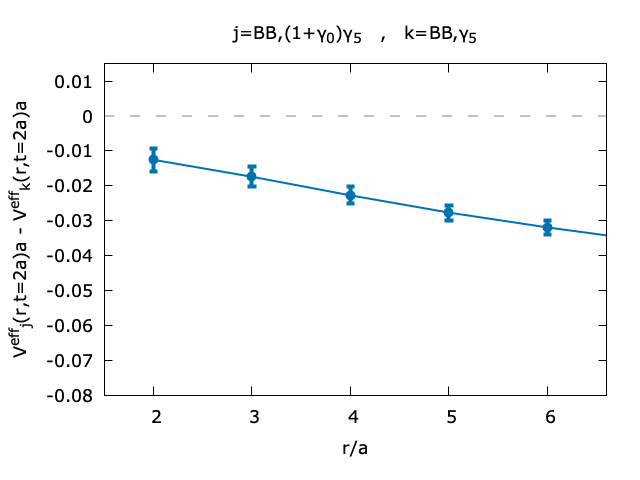}
\hfill
\includegraphics[width=0.49\columnwidth]{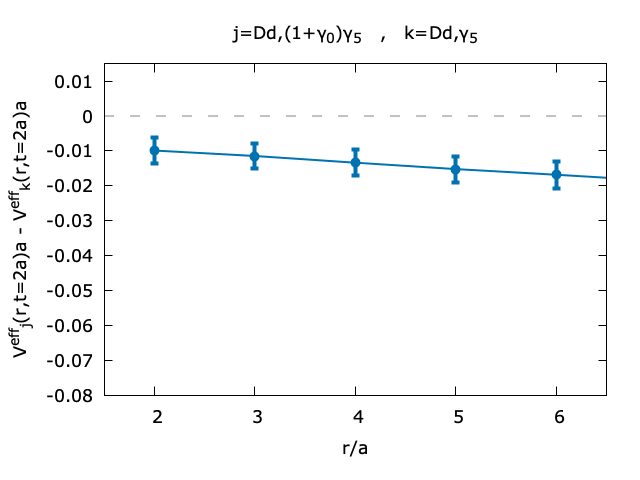} \\
\includegraphics[width=0.49\columnwidth]{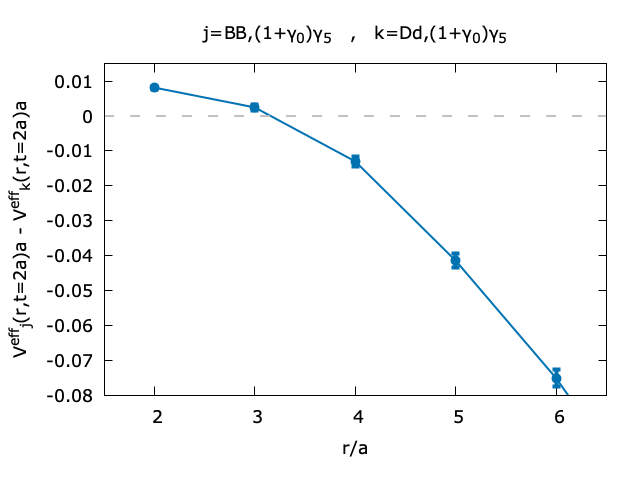}
\caption{\label{fig:meff}$V_j^\text{eff}(r,t=2a) - V_k^\text{eff}(r,t=2a)$, i.e.\ differences of effective energies corresponding to diagonal elements of the correlation matrix for $t/a = 2$, as functions of $r$ for ensemble B40.24. \textbf{(top left)}~$j = BB , (1+\gamma_0) \gamma_5$, $k = BB , \gamma_5$. \textbf{(top right)}~$j = Dd , (1+\gamma_0) \gamma_5$, $k = Dd , \gamma_5$. \textbf{(bottom)}~$j = BB , (1+\gamma_0) \gamma_5$, $k = Dd , (1+\gamma_0) \gamma_5$.}
\end{figure}

We start with a comparison of the two operators $\mathcal{O}_{BB,(1 + \gamma_0) \gamma_5}$ and $\mathcal{O}_{BB,\gamma_5}$ for ensemble B40.24 by showing the difference of their effective energies, $V_{BB,(1 + \gamma_0) \gamma_5}^\text{eff}(r, t=2a) - V_{BB,\gamma_5}^\text{eff}(r, t=2a)$, in the upper left plot of Figure~\ref{fig:meff}. This difference is clearly negative for all separations $r$, which is not surprising. $\mathcal{O}_{BB,(1 + \gamma_0) \gamma_5}$ creates predominantly a pair of ground state static-light mesons, while $\mathcal{O}_{BB,\gamma_5}$ creates roughly a $50\% / 50\%$ superposition of a pair of negative parity ground state mesons and a pair of significantly heavier positive parity static-light mesons, as discussed above and as can be shown e.g.\ by a Fierz transformation (for details see Ref.\ \cite{Bicudo:2015kna}). Results from an analog comparison of the two diquark-antidiquark operators are very similar (see upper right plot of Figure~\ref{fig:meff}). This is interesting, because it indicates that a light diquark with spin structure given by $(1 + \gamma_0) \gamma_5$ is energetically preferred over a light diquark with spin structure given just by $\gamma_5$.

% B: 3.15 * a = 3.15 * 0.079 fm = 0.25 fm
% C: 3.75 * a = 3.65 * 0.063 fm = 0.23 fm

Most interesting, of course, is the comparison of a meson-meson and a diquark-antidiquark operator, specifically of $\mathcal{O}_{BB,(1 + \gamma_0) \gamma_5}$ and $\mathcal{O}_{Dd,(1 + \gamma_0) \gamma_5}$, which we have just identified as being superior to $\mathcal{O}_{BB,\gamma_5}$ and $\mathcal{O}_{Dd,\gamma_5}$, respectively. We show the difference of their effective masses, $V_{BB,(1 + \gamma_0) \gamma_5}^\text{eff}(r, t=2a) - V_{Dd,(1 + \gamma_0) \gamma_5}^\text{eff}(r, t=2a)$, in the lower plot of Figure~\ref{fig:meff}. For $r \ltapprox 3.15 \, a \approx 0.25 \, \text{fm}$ the difference is positive, indicating that for small separations the diquark-antidiquark operator generates a trial state more similar to the ground state. For larger separations, $r \gtapprox 0.25 \, \text{fm}$ the difference becomes negative and strongly points towards a meson-meson structure. While the latter is expected (at large $r$ the flux tube present in a heavy diquark-antidiquark is energetically disfavored), the former is a first hint towards a diquark-antidiquark dominance at smaller $r$. We continue to investigate this in more detail in the following subsections.

Analog plots of differences of effective energies for ensemble C30.32 are very similar to those of Figure~\ref{fig:meff} and, thus, not shown.

% **********

\subsection{\label{SEC495}Optimizing trial states by minimizing effective energies}

Now we consider the 2-dimensional space spanned by the states $| \Phi_{BB , (1+\gamma_0) \gamma_5} \rangle$ and $| \Phi_{Dd , (1+\gamma_0) \gamma_5} \rangle$. Any trial state from that space can be written as
\begin{eqnarray}
\label{EQN643} | \Phi_{b,d} \rangle = b | \Phi_{BB , (1+\gamma_0) \gamma_5} \rangle + d | \Phi_{Dd , (1+\gamma_0) \gamma_5} \rangle
\end{eqnarray}
with coefficients $b, d \in \mathbb{C}$. To identify a trial state as similar to the ground state as possible, i.e.\ with large overlap to the ground state and little overlap to excitations, we minimize the corresponding effective energy
\begin{eqnarray}
\label{EQN423} V_{b,d}^\text{eff}(r,t) = -\frac{1}{a} \log\bigg(\frac{C_{[b,d] [b,d]}(t)}{C_{[b,d] [b,d]}(t-a)}\bigg)
\end{eqnarray}
with respect to $b$ and $d$. Since, $V_{b,d}^\text{eff}$ is independent of the norm and the phase of $| \Phi_{b,d} \rangle$, we can fix $b = 1$ and minimize $V_{b,d}^\text{eff}$ for given $r$ and $t$ with respect to the real and the imaginary part of $d$. Such a 2-dimensional minimization is numerically straightforward. In particular, no further lattice QCD computations are needed, because $C_{[b,d] [b,d]}$ can be expressed in terms of the correlation matrix introduced in Eq.\ (\ref{EQN755}),
\begin{eqnarray}
C_{[b,d] [b,d]}(t) =
\left(\begin{array}{c} b \\ d \end{array}\right)^\dagger_j
C_{j k}(t)
\left(\begin{array}{c} b \\ d \end{array}\right)_k .
\end{eqnarray}

\begin{figure}[htb]
\includegraphics[width=0.28\columnwidth]{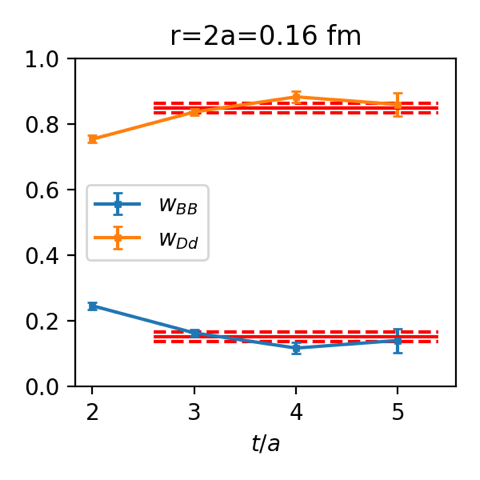}
\includegraphics[width=0.28\columnwidth]{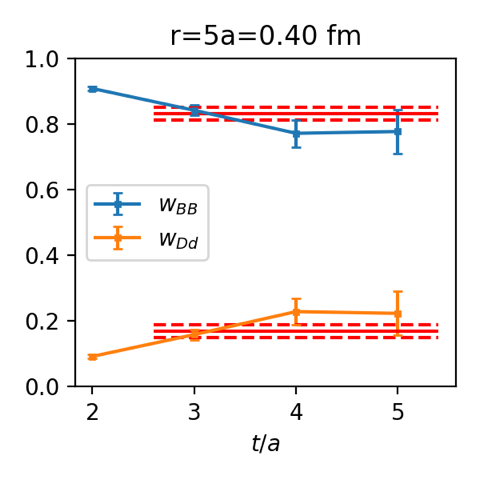}
\includegraphics[width=0.28\columnwidth]{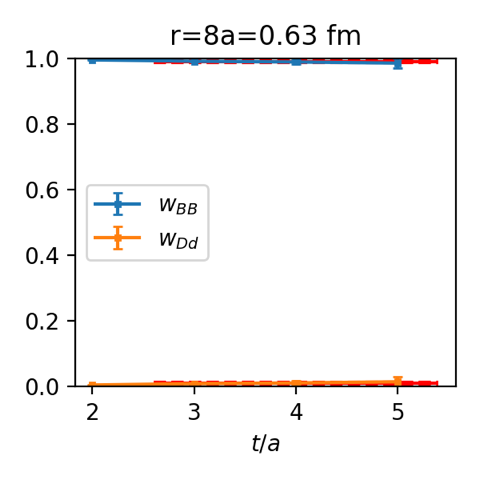}
\caption{\label{FIG600}$w_{BB}$ and $w_{Dd} = 1 - w_{BB}$, the normalized absolute squares of the coefficients of the optimized trial states for several fixed $r$ as functions of $t$ for ensemble B40.24. The horizontal red lines indicate the fit results $\bar{w}_{BB}$ and $\bar{w}_{Dd}$ and the corresponding statistical errors.}
\end{figure}

In the following we consider
\begin{eqnarray}
w_{BB} = \frac{|b|^2}{|b|^2 + |d|^2} \quad , \quad w_{Dd} = \frac{|d|^2}{|b|^2 + |d|^2} = 1 - w_{BB}
\end{eqnarray}
with $b = 1$ and $d$ minimizing $V_{b,d}^\text{eff}$. $w_{BB}$ and $w_{Dd}$ are the normalized absolute squares of the coefficients of the optimized trial states appearing in Eq.\ (\ref{EQN643}). These quantities exhibit only a weak dependence on $t$. For $3 \leq t/a \leq 5$ and ensemble B40.24 ($4 \leq t/a \leq 6$ and ensemble C30.32) they are consistent with a constant. For $t/a \geq 6$ ($t/a \geq 7$), statistical fluctuations and errors become large and the signal is quickly lost in noise. The latter is not surprising, because $w_{BB}$ and $w_{Dd}$ are subtle quantities depending on the amount of excited states in $BB$ and $Dd$ correlation functions, which are exponentially suppressed in $t$. In Figure~\ref{FIG600} we show example plots of $w_{BB}$ and $w_{Dd}$ as functions of $t$ for selected separations $r/a = 2$, $r/a = 5$ and $r/a = 8$ for ensemble~B.

\begin{figure}[htb]
\includegraphics[width=0.49\columnwidth]{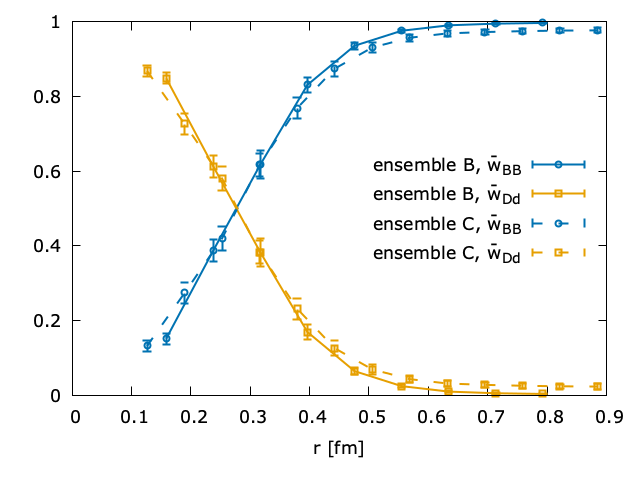}
\caption{\label{FIG601}$\bar{w}_{BB}$ and $\bar{w}_{Dd} = 1 - \bar{w}_{BB}$, the fitted normalized absolute squares of the coefficients of the optimized trial state, as functions of $r$ for both ensembles.}
\end{figure}

We determine each plateau value by a $\chi^2$ minimizing fit of a constant in the range $3 \leq t/a \leq 5$ ($4 \leq t/a \leq 6$). The resulting numbers, $\bar{w}_{BB}(r)$ and $\bar{w}_{Dd}(r) = 1 - \bar{w}_{BB}(r)$, can be interpreted as the relative weight of a meson-meson and a diquark-antidiquark structure at $\bar{b} \bar{b}$ separation $r$ in the ground state, which corresponds to the potential $V(r)$ of two static antiquarks and is, thus, closely related to the $\bar{b} \bar{b} u d$ tetraquark with quantum numbers $I(J^P) = 0(1^+)$. In Figure~\ref{FIG601} we plot $\bar{w}_{BB}$ and $\bar{w}_{Dd}$ as functions of $r$. One can clearly see that there is a diquark-antidiquark dominance for $\bar{b} \bar{b}$ separations $r \ltapprox 0.20 \, \textrm{fm}$. For $r \gtapprox 0.30 \, \textrm{fm}$, the meson-meson structure is more prominent and for $r \gtapprox 0.50 \, \textrm{fm}$, the diquark-antidiquark contribution is negligible, i.e.\ the system is exclusively composed of two $B$ mesons. It is interesting to note that the separation $r \approx 0.3 \, \text{fm}$, where the meson-meson structure starts to dominate, is of the same order as the size of a $B$ meson. A precise comparison, however, seems to be difficult, because the size of a $B$ meson is model dependent and not precisely known (see e.g.\ Refs.\ \cite{Hwang:2001th,Das:2016rio}).

We note that the quantities $w_{BB}$ and $w_{Dd}$ as well as the fitted $\bar{w}_{BB}$ and $\bar{w}_{Dd}$ depend on the normalization of the operators $\mathcal{O}_{BB,(1 + \gamma_0) \gamma_5}$ and $\mathcal{O}_{Dd,(1 + \gamma_0) \gamma_5}$ or, equivalently, on the normalization and the corresponding states $| \Phi_{BB , (1+\gamma_0) \gamma_5} \rangle$ and $| \Phi_{Dd , (1+\gamma_0) \gamma_5} \rangle$. To allow a meaningful interpretation in terms of the relative weight of a $BB$ and a $Dd$ structure, the norms of these two states has to be similar. In this work we use $N_{BB} = N_{Dd}$, which implies $\mathcal{O}_{BB,\Gamma} = \mathcal{O}_{Dd,\Gamma}$ for $r = 0$ and, thus, $| \Phi_{BB,\Gamma} \rangle = | \Phi_{Dd,\Gamma} \rangle$ for $r = 0$ (see section~\ref{sec:lattice}). We expect that $N_{BB} = N_{Dd}$ also results in similar norms for $r > 0$. A common alternative, which we have used in previous lattice QCD projects, is to normalize the operators $\mathcal{O}_j$ such that $C_{jj}(t=a) = 1$ (no sum over $j$; see e.g.\ Refs.\ \cite{Kalinowski:2015bwa,Alexandrou:2017itd,Alexandrou:2019tmk}). As a cross-check we also explored this normalization in our current work and found almost identical results to those obtained with $N_{BB} = N_{Dd}$.

% **********

\subsection{\label{SEC607}Eigenvector components obtained by solving a generalized eigenvalue problem}

To further explore the structure of the ground state in the $(I,j_z,\mathcal{P},\mathcal{P}_x) = (0,0,-,+)$ sector, we now use $N \times N$ correlation matrices $C_{jk}$ as defined in Eq.\ (\ref{EQN755}) and solve the generalized eigenvalue problem (GEVP)
\begin{eqnarray}
\label{EQN244} C_{jk}(t) v_k^{(n)}(t) = \lambda^{(n)}(t) C_{jk}(t_0) v_k^{(n)}(t) \quad, \quad n = 0,\dots,N-1
\end{eqnarray}
for $t_0/a \geq 1$ and $t/a > t_0/a$ (for detailed discussions of the GEVP in lattice field theory see e.g.\ Refs.\ \cite{Luscher:1990ck,Danzer:2007bx,Blossier:2009kd,Bulava:2011np,Shultz:2015pfa,Schiel:2015kwa,Dragos:2016rtx,Fischer:2020bgv}). Effective energies for the lowest $N$ energy eigenstates are then given by
\begin{eqnarray}
\label{EQN697} V^{\text{eff},(n)}(r,t) = -\frac{1}{a} \log\bigg(\frac{\lambda^{(n)}(t)}{\lambda^{(n)}(t-a)}\bigg) .
\end{eqnarray}
These are generalizations of the effective energy defined in Eq.\ (\ref{EQN539}), since $V^{\text{eff},(0)}(r,t) = V_j^{\text{eff}}(r,t)$ for $N = 1$. For $N > 1$, $V^{\text{eff},(0)}(r,t)$ approaches the same constant $V(r)$ for large $t$, but plateaus can typically be identified at somewhat smaller $t$, because of an elimination of excitations (see also the second next paragraph and appendix~\ref{APP100}, where a minimization of effective energies is related to the GEVP).

The eigenvector components $v_j^{(n)}(t)$, which we always normalize according to $\sum_j |v_j^{(n)}(t)|^2 = 1$, contain information about the relative importance of the creation operators included in the correlation matrix and, thus, hints about the structure of the corresponding energy eigenstates. For large $t$ and $t_0$,
\begin{eqnarray}
| n \rangle \approx \sum_j v_j^{(n)}(t) | \Phi_j \rangle ,
\end{eqnarray}
where the $\approx$ sign denotes an approximate expansion of the energy eigenstate $| n \rangle$ in terms of the trial states $| \Phi_j \rangle$. For such values of $t$ and $t_0$, the squared eigenvector components as functions of $t$ form plateaus and we determine the corresponding asymptotic values of $|v_j^{(0)}(t)|^2$ by $\chi^2$ minimizing fits of constants. The results of these fits are denoted by $|\bar{v}_j^{(0)}|^2$. The squared eigenvector components $|v_j^{(n)}(t)|^2$ as well as the fitted $|\bar{v}_j^{(0)}|^2$ depend on the normalization of the creation operators, as it is the case for $w_{BB}$, $w_{Dd}$, $\bar{w}_{BB}$ and $\bar{w}_{Dd}$ (see the discussion at the end of section~\ref{SEC495}). As before, we use $N_{BB} = N_{Dd}$, which amounts to having trial states $| \Phi_j \rangle$ with similar norm.

It is interesting to note that for a $2 \times 2$ correlation matrix with trial states $| \Phi_1 \rangle = | \Phi_{BB , (1+\gamma_0) \gamma_5} \rangle$ and \\ $| \Phi_2 \rangle = | \Phi_{Dd , (1+\gamma_0) \gamma_5} \rangle$ and $t_0/a = t/a - 1$, the eigenvector components $v_{j}^{(0)}$ are proportional to the coefficients $b$ and $d$ minimizing $V_{b,d}^\text{eff}(r,t)$ defined in Eq.\ (\ref{EQN423}). Moreover, $(|v_{BB , (1+\gamma_0) \gamma_5}^{(0)}|^2 , |v_{Dd , (1+\gamma_0) \gamma_5}^{(0)}|^2) = (w_{BB} , w_{Dd})$ and, thus, \\ $(|\bar{v}_{BB , (1+\gamma_0) \gamma_5}^{(0)}|^2 , |\bar{v}_{Dd , (1+\gamma_0) \gamma_5}^{(0)}|^2) = (\bar{w}_{BB} , \bar{w}_{Dd})$, as we show in appendix~\ref{APP100}. In other words, the results on the structure of the $\bar{b} \bar{b} u d$ ground state obtained in section~\ref{SEC495} by determining a trial state, which minimizes an effective energy, are identical to results from a specific corresponding GEVP. This allows to understand and interpret the GEVP eigenvector components from another perspective. Compared to the trial state optimization from section~\ref{SEC495}, the GEVP, however, offers further possibilities, for example to choose $t_0$ independent of $t$ (i.e.\ not as $t_0/a = t/a - 1$) or to study the full correlation matrix with $N = 4$.

\begin{figure}[htb]
\includegraphics[width=0.28\columnwidth]{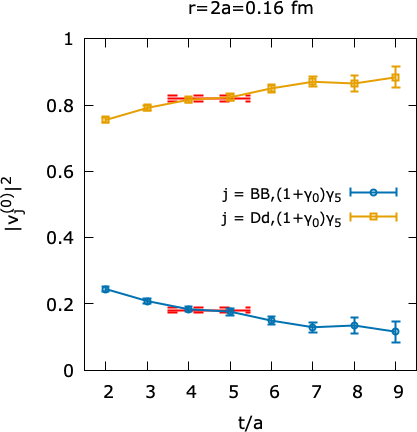}
\includegraphics[width=0.28\columnwidth]{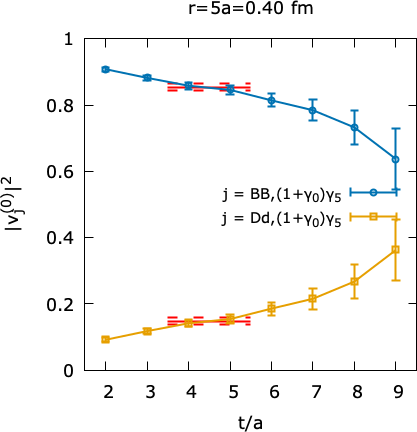}
\includegraphics[width=0.28\columnwidth]{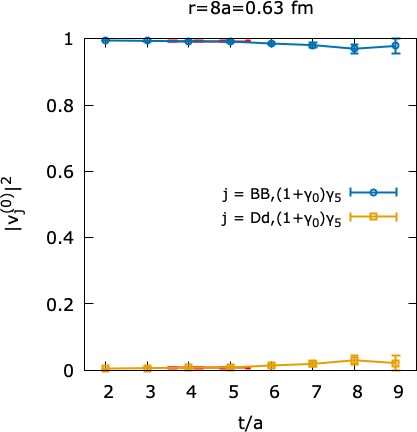} \\
\includegraphics[width=0.28\columnwidth]{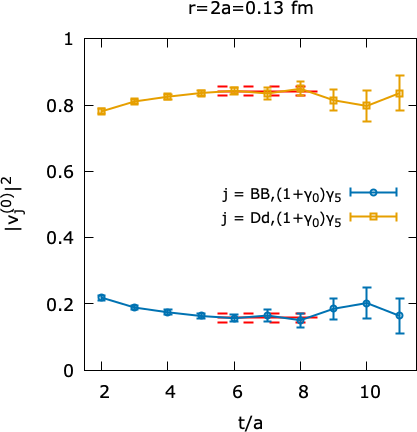}
\includegraphics[width=0.28\columnwidth]{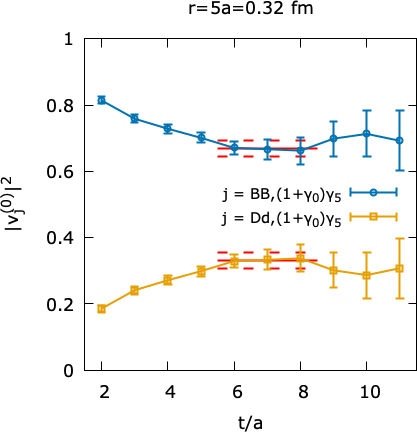}
\includegraphics[width=0.28\columnwidth]{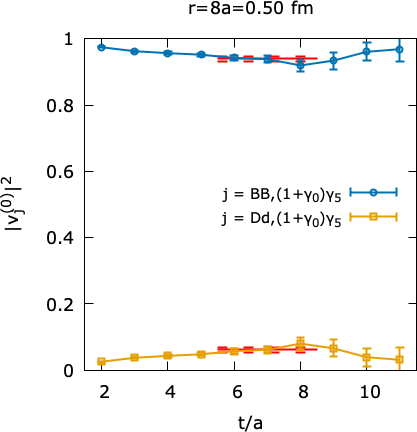}
\caption{\label{FIG602}The squared eigenvector components $|v_{BB , (1+\gamma_0) \gamma_5}^{(0)}|^2$ and $|v_{Dd , (1+\gamma_0) \gamma_5}^{(0)}|^2 = 1 - |v_{BB , (1+\gamma_0) \gamma_5}^{(0)}|^2$ for several fixed $r$ as functions of $t$. \textbf{(upper line)}~Ensemble B40.24. \textbf{(lower line)}~Ensemble C30.32. The horizontal red lines indicate the fit results $|\bar{v}_{BB , (1+\gamma_0) \gamma_5}^{(0)}|^2$ and $|\bar{v}_{Dd , (1+\gamma_0) \gamma_5}^{(0)}|^2$ and the corresponding statistical errors.}
\end{figure}

As discussed in the previous paragraph, solving the GEVP with a $2 \times 2$ correlation matrix including the operators $\mathcal{O}_{BB,(1 + \gamma_0) \gamma_5}$ and $\mathcal{O}_{Dd,(1 + \gamma_0) \gamma_5}$ and using $t_0/a = t/a - 1$ yields exactly the same results as shown in Figure~\ref{FIG601} (one just has to replace labels according to $\bar{w}_{BB} \rightarrow |\bar{v}_{BB,(1 + \gamma_0) \gamma_5}^{(0)}|^2$ and $\bar{w}_{Dd} \rightarrow |\bar{v}_{Dd,(1 + \gamma_0) \gamma_5}^{(0)}|^2$). However, many lattice QCD papers using the GEVP fix $t_0$ to a rather small value independent of $t$. For small values of $t_0$, statistical errors are somewhat reduced, which in turn allows to consider larger values of $t$. Thus we also computed $|\bar{v}_{BB , (1+\gamma_0) \gamma_5}^{(0)}|^2$ and $|\bar{v}_{Dd , (1+\gamma_0) \gamma_5}^{(0)}|^2$ for $t_0/a = 1$. As already observed in the previous subsection for the related quantities $w_{BB}$ and $w_{Dd}$, the resulting squared eigenvector components $|v_{BB , (1+\gamma_0) \gamma_5}^{(0)}|^2$ and $|v_{Dd , (1+\gamma_0) \gamma_5}^{(0)}|^2$ exhibit only a mild $t$ dependence and the majority is consistent with a constant for large $t$. Results for selected $\bar{b} \bar{b}$ separations $r$ are shown in Figure~\ref{FIG602}. For ensemble C30.32 we use the fit range $6 \leq t/a \leq 8$ to determine $|\bar{v}_{BB , (1+\gamma_0) \gamma_5}^{(0)}|^2$ and $|\bar{v}_{Dd , (1+\gamma_0) \gamma_5}^{(0)}|^2$. For ensemble B40.24 the analysis and interpretation of the data is less obvious. While for $t/a \leq 5$ plateaus are indicated, for $t/a \geq 6$ there is a weak, almost linearly increasing deviation from these plateaus for some separations $r$ (see e.g.\ the plot in the center of the upper line of Figure~\ref{FIG602}). Note, however, that errors are also increasing at larger $t$. For example, the data points at $8 \leq t/a \leq 9$ are just around $2 \sigma$ away from those at $4 \leq t/a \leq 5$. Also the monotonic almost linear behavior is not necessarily an indication of a systematic deviation from a constant, because all data points were computed on the same gauge link configurations and neighboring points in $t$ are, thus, correlated. Consequently, we interpret the observed deviations as statistical fluctuations and use the fit range $4 \leq t/a \leq 5$.

\begin{figure}[htb]
\includegraphics[width=0.49\columnwidth]{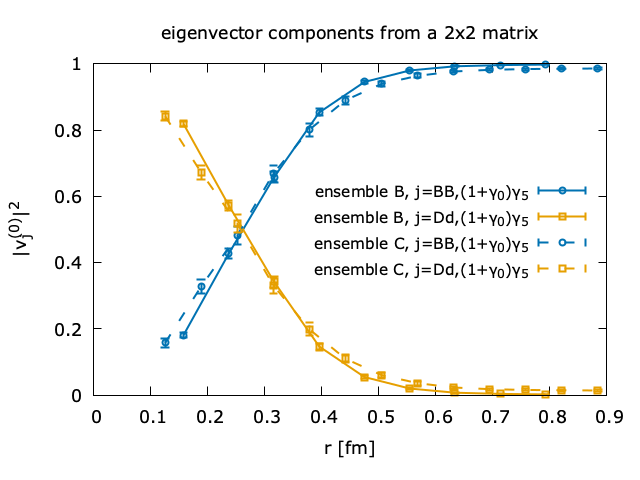}
\hfill
\includegraphics[width=0.49\columnwidth]{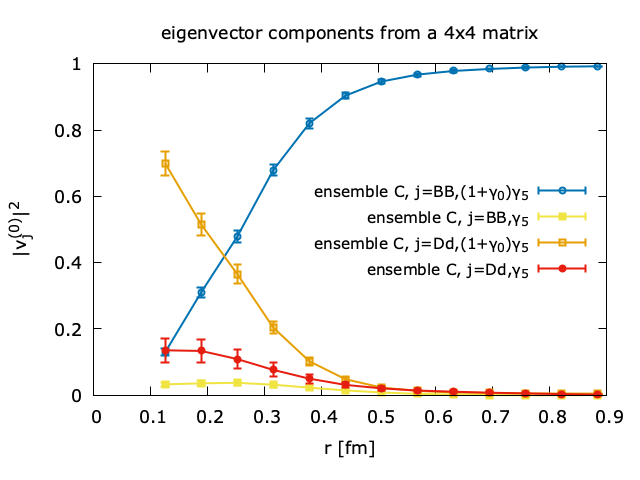}
\caption{\label{fig:v_sq}Fitted squared eigenvector components $|\bar{v}_j^{(0)}|^2$ as functions of $r$. \textbf{(left)}~$2 \times 2$ correlation matrix including the operators $\mathcal{O}_{BB,(1 + \gamma_0) \gamma_5}$ and $\mathcal{O}_{Dd,(1 + \gamma_0) \gamma_5}$ for both ensembles. \textbf{(right)}~$4 \times 4$ correlation matrix including all operators (see Eq.\ (\ref{EQN868})) for ensemble C30.32.}
\end{figure}

In the left plot of Figure~\ref{fig:v_sq} we show $|\bar{v}_{BB , (1+\gamma_0) \gamma_5}^{(0)}|^2$ and $|\bar{v}_{Dd , (1+\gamma_0) \gamma_5}^{(0)}|^2$ for both ensembles. These curves for $t_0/a = 1$ are quite similar to those corresponding to $t_0/a = t/a - 1$ (and shown in Figure~\ref{FIG601}). Again, one can see a clear dominance of the diquark-antidiquark operator for separations $r \ltapprox 0.20 \, \text{fm}$. In the range $0.20 \, \text{fm} \leq r \leq 0.30 \, \text{fm}$ there is a rapid change towards a meson-meson structure. For $r \gtapprox 0.50 \, \text{fm}$ there is almost no diquark-antidiquark contribution anymore and the $\bar{b} \bar{b} u d$ four-quark system seems to be composed exclusively of two $B$ mesons. This plot confirms our results from section~\ref{SEC494} obtained by minimizing effective energies.

It should be noted that the quantities $\bar{w}_{BB}$ and $\bar{w}_{Dd}$ as well as the fitted squared eigenvector components $|\bar{v}_{BB , (1+\gamma_0) \gamma_5}^{(0)}|^2$ and $|\bar{v}_{Dd , (1+\gamma_0) \gamma_5}^{(0)}|^2$ depend on the creation operators used, e.g.\ the details of the quark and gauge field smearing or their normalization. Besides discretization errors, this might be part of the reason for the slight differences between the results obtained for ensemble B40.24 and ensemble C30.32 shown in Figure~\ref{FIG601} and the left plot of Figure~\ref{fig:v_sq}. Since these differences are quite small and since we found almost identical results for different normalization prescriptions, we consider $\bar{w}_{BB}$ and $\bar{w}_{Dd}$ as well as $|\bar{v}_j^{(0)}|^2$ as reliable indicators characterizing the quark and gluon structure of the ground state of the $(I,j_z,\mathcal{P},\mathcal{P}_x) = (0,0,-,+)$ sector.

Finally we performed the same GEVP analysis using the full $4 \times 4$ correlation matrix including all operators defined in Eq.\ (\ref{EQN868}). We determined $|\bar{v}_j^{(0)}|^2$ using the same fit ranges as for the previous $2 \times 2$ analyses (results for ensemble C30.32 are shown in the right plot of Figure~\ref{fig:v_sq}). Again, there is a diquark-antidiquark dominance for $r \ltapprox 0.20 \, \text{fm}$, this time represented by the eigenvector components corresponding to two operators $\mathcal{O}_{Dd,(1 + \gamma_0) \gamma_5}$ and $\mathcal{O}_{Dd,\gamma_5}$, while there is meson-meson dominance for $r \gtapprox 0.30 \, \text{fm}$, reflected by the eigenvector components corresponding to two operators $\mathcal{O}_{BB,(1 + \gamma_0) \gamma_5}$ and $\mathcal{O}_{BB,\gamma_5}$. When adding the two diquark eigenvector components as well as the two meson-meson eigenvector components, i.e.\ when considering $|\bar{v}_{Dd,(1 + \gamma_0) \gamma_5}^{(0)}|^2 + |\bar{v}_{Dd,\gamma_5}^{(0)}|^2$ and $|\bar{v}_{BB,(1 + \gamma_0) \gamma_5}^{(0)}|^2 + |\bar{v}_{BB,\gamma_5}^{(0)}|^2$, we find within errors the same curves as obtained above by the $2 \times 2$ analysis. This is reassuring, because the results concerning the meson-meson percentage and the diquark-antidiquark percentage do not change, even though we increased the dimension of the basis of trial states used to approximate the ground state from $2$ to $4$.

We note that the GEVP can also be used to study excited states. In our case, i.e.\ for quantum numbers $(I,j_z,\mathcal{P},\mathcal{P}_x) = (0,0,-,+)$, the first and second excitation correspond to the repulsive potential of a negative and a positive parity $B$ meson and to the attractive potential of two positive parity $B$ mesons. We computed these potentials in a pevious work \cite{Bicudo:2015kna} (see in particular Figure~4 in Ref.\ \cite{Bicudo:2015kna}, ``singlet~A''), where operators $\mathcal{O}_{BB,\Gamma}$ with a larger set of matrices $\Gamma$ were used. To study the structure of these excitations would require also a comparable set of operators $\mathcal{O}_{Dd,\Gamma}$ and goes beyond the scope of this work.

% **********

\subsection{\label{SEC600}Meson-meson and diquark-antidiquark percentages of the $\bar{b} \bar{b} u d$ tetraquark with $I(J^P) = 0(1^+)$}

During the lattice QCD computation of the potential $V(r)$ the heavy antiquarks $\bar{b} \bar{b}$ are considered as static, i.e.\ their positions are fixed and only the light quarks $u d$ and the gluons are dynamical degrees of freedom. The dynamics of the heavy quarks can, however, be studied in a second step, by inserting the potential $V(r)$ into the Schr\"odinger equation (\ref{EQN649}). This two step approach is widely known as the Born-Oppenheimer approximation (see e.g.\ Ref.\ \cite{Braaten:2014qka} for a detailed discussion in the context of exotic mesons). In Ref.\ \cite{Bicudo:2012qt} we solved the Schr\"odinger equation (\ref{EQN649}) using $m_b = m_B = 5279 \, \text{MeV}$ \cite{Zyla:2020zbs} and found a single bound state with binding energy $-E = 38(18) \, \text{MeV}$ indicating the existence of a hadronically stable $\bar{b} \bar{b} u d$ tetraquark with quantum numbers $I(J^P) = 0(1^+)$. Moreover, the wave function $R(r) / r$ gives the probability density of the $\bar{b} \bar{b}$ separation, $p_r(r) = 4 \pi |R(r)|^2$, which is shown in the right plot of Figure~\ref{FIG900}.

The quantities $\bar{w}_{BB}$ and $\bar{w}_{Dd}$ (see Figure~\ref{FIG601}) as well as $|\bar{v}_{BB,(1 + \gamma_0) \gamma_5}^{(0)}|^2$ and $|\bar{v}_{Dd,(1 + \gamma_0) \gamma_5}^{(0)}|^2$ (see left plot of Figure~\ref{fig:v_sq}) can be interpreted as meson-meson and diquark-antidiquark percentages for fixed $r$. We define
\begin{eqnarray}
p_{BB,w}(r) = \bar{w}_{BB} \quad , \quad p_{Dd,w}(r) = \bar{w}_{Dd}
\end{eqnarray}
and
\begin{eqnarray}
p_{BB,v}(r) = |\bar{v}_{BB,(1 + \gamma_0) \gamma_5}^{(0)}|^2 \quad , \quad p_{Dd,v}(r) = |\bar{v}_{Dd,(1 + \gamma_0) \gamma_5}^{(0)}|^2 ,
\end{eqnarray}
where $p_{BB,w}(r) \approx p_{BB,v}(r)$ and $p_{Dd,w}(r) \approx p_{Dd,v}(r)$. These percentages together with the probabilty density $p_r(r)$ can be used to crudely estimate the total meson-meson and diquark-antidiquark percentages of the $\bar{b} \bar{b} u d$ tetraquark, which we denote by $\%BB_j$ and $\%Dd_j$. To this end, we use a parameterization, which is a simple mathematical function able to consistently describe our lattice QCD results,
\begin{eqnarray}
\label{EQN422} p_{BB,j}(r) = \frac{1}{2} \Big(\textrm{tanh} (\alpha_j (r-r_{0,j}) + 1\Big) \quad , \quad p_{Dd,j}(r) = 1 - p_{BB,j}(r) ,
\end{eqnarray}
where the parameters $r_{0,j}$ and $\alpha_j$ are determined by $\chi^2$ minimizing fits to the results of both ensembles as shown in Figure~\ref{FIG601} and Figure~\ref{fig:v_sq}. Then we compute $\%BB$ and $\%Dd$ via
\begin{eqnarray}
\label{EQN424} \%BB_j = \int dr \, p_r(r) p_{BB,j}(r) \quad , \quad \%Dd_j = \int dr \, p_r(r) p_{Dd,j}(r) = 1 - \%BB_j .
\end{eqnarray}
We find $\%BB_w = 0.58$, $\%Dd_w = 0.42$ and $\%BB_v = 0.60$, $\%Dd_v = 0.40$, i.e.\ almost the same result, when using $\bar{w}_{BB}$ and $\bar{w}_{Dd}$ and when using $|\bar{v}_{BB,(1 + \gamma_0) \gamma_5}^{(0)}|^2$ and $|\bar{v}_{Dd,(1 + \gamma_0) \gamma_5}^{(0)}|^2$. Note that we have lattice QCD results for $p_{BB,j}(r)$ and $p_{Dd,j}(r)$ only for separations $r \gtapprox 0.1 \, \text{fm}$. Thus, it is unclear, whether the parameterization (\ref{EQN422}) is a valid description also for $r \ltapprox 0.1 \, \text{fm}$. The coresponding systematic error is, however, quite small, because the probability to find the $\bar{b} \bar{b}$ pair at separation $r \ltapprox 0.1 \, \text{fm}$ is also rather small (see the plot of $p_r(r) = 4 \pi |R(r)|^2$ in Figure~\ref{FIG900}). To quote a crude and very conservative upper bound for this systematic error, we solved again the integral in Eq.\ (\ref{EQN424}) replacing in the interval $0 \leq r \leq 0.1 \, \text{fm}$ the almost vanishing $p_{BB,j}(r) = (\textrm{tanh} (\alpha_j (r-r_{0,j}) + 1) / 2 \approx 0$ by $p_{BB,j}(r) = 1$. The results are quite similar, $\%BB_w = 0.63$, $\%Dd_w = 0.37$ and $\%BB_v = 0.65$, $\%Dd_v = 0.35$, indicating that the corresponding systematic error is well below $0.05$. Moreover, the eigenvector components are slightly operator dependent, as discussed in section~\ref{SEC607}. Thus, the percentages $\%BB_j$ and $\%Dd_j$ should only be considered as crude estimates. As total systematic error, reflecting both the parameterization and the operator dependence, we estimate $\approx 0.10$. Still it seems to be clear that the $\bar{b} \bar{b} u d$ tetraquark with quantum numbers $I(J^P) = 0(1^+)$ is neither strongly meson-meson dominated nor strongly diquark-antidiquark dominated, but rather an approximately equal linear combination of both structures. Finally we note that these results for $\%BB_j$ and for $\%Dd_j$ are fully consistent with squared eigenvector components obtained during a recent lattice QCD study \cite{Leskovec:2019ioa} of the same tetraquark using four quarks of finite mass. There, a meson-meson component of $0.65(4)$ and a diquark-antidiquark component of $0.35(4)$ was found \cite{Pflaumer2020}.

%%%%%%%%%%%%%%%%%%%%%%%%%%%%%%%%%%%%%%%%%%%%%%%%%%%%%%%%%%%%%%%%%%%%%%%
%%
%%                                   SSS                             
%%                                  S   S                            
%%                                   S                               
%%                                    S                              
%%                                     S                             
%%                                  S   S                            
%%                                   SSS                             
%%
%%%%%%%%%%%%%%%%%%%%%%%%%%%%%%%%%%%%%%%%%%%%%%%%%%%%%%%%%%%%%%%%%%%%%%%

% \newpage

\section{Conclusions and outlook \label{sec:conclusion}}

In this work we used lattice QCD to study a recurrent question on the nature of tetraquarks in the context of a $\bar{b} \bar{b} u d$ tetraquark with quantum numbers $I(J^P) = 0(1^+)$: Are they more similar to meson-meson systems or rather to diquark-antidiquark pairs? Moreover we addressed the Dirac structure of the light quarks, comparing $\gamma^5$ with $(1+ \gamma^0) \gamma^5$, which are both consistent with $I(J^P) = 0(1^+)$.

We implemented four different lattice QCD creation operators, two of meson-meson type and two of diquark-antidiquark type. We solved the GEVP both for $2 \times 2$ and $4 \times 4$ correlation matrices, to determine quantitatively, which of the implemented structures is preponderant in the ground state at $\bar b \bar b$ separation $r$. Moreover, we optimized trial states by minimizing effective energies and proved that this is equivalent to solving a specific corresponding GEVP.

Notice that the question we are addressing is quite subtle. We first showed that the $BB$ and $Dd$ trial states are not orthogonal. Nevertheless, since they are not linearly dependent either, we were able to determine, which one is more similar to the ground state. In what concerns light spin we found the $(1+ \gamma^0) \gamma^5$ is the dominant Dirac structure at all separations $r$, both for $BB$ and for $Dd$. This is what we expected, since the less favorable $\gamma^5$ structure generates not only negative parity mesons, but also excited positive parity mesons. As for color we showed that at small separations $r \ltapprox 0.25 \, \text{fm}$ the diquark-antidiquark structure dominates, whereas at larger separations the meson-meson trial state is clearly more similar to the ground state of the $\bar{b} \bar{b} u d$ system. For $r \gtapprox 0.5 \, \text{fm}$, the percentage of $BB$ is already larger than $95 \%$ and approaches $100 \%$ for even larger $r$.

Using these results as well as the wave function of the $\bar{b} \bar{b}$ separation already obtained in Ref.\ \cite{Bicudo:2012qt}, we estimated the meson-meson to diquark-antidiquark ratio of the $\bar{b} \bar{b} u d$ tetraquark with quantum numbers $I(J^P) = 0(1^+)$ and found around $60 \% / 40 \%$. Thus $BB$ and $Dd$ components seem to be present in comparable parts.

%%%%%%%%%%%%%%%%%%%%%%%%%%%%%%%%%%%%%%%%%%%%%%%%%%%%%%%%%%%%%%%%%%%%%%%
%%
%%                                   SSS                             
%%                                  S   S                            
%%                                   S                               
%%                                    S                              
%%                                     S                             
%%                                  S   S                            
%%                                   SSS                             
%%
%%%%%%%%%%%%%%%%%%%%%%%%%%%%%%%%%%%%%%%%%%%%%%%%%%%%%%%%%%%%%%%%%%%%%%%

% \newpage

\appendix

\section{\label{APP100}Equivalence of optimizing a trial state by minimizing an effective energy and of solving a GEVP}

We start with the $N \times N$ GEVP defined in Eq.\ (\ref{EQN244}), choose $t_0 = t-a$ and use the simplified notation \\ $\mathbf{v}^{(n)} \equiv (v_1^{(n)}(t) , v_2^{(n)}(t) , \ldots , v_N^{(n)}(t))$ and $\lambda^{(n)} \equiv \lambda^{(n)}(t)$,
\begin{eqnarray}
\label{EQN670} C(t) \mathbf{v}^{(n)} = \lambda^{(n)} C(t-a) \mathbf{v}^{(n)} \quad, \quad n = 0,\dots,N-1 .
\end{eqnarray}
The correlation matrix $C$ defined in Eq.\ (\ref{EQN755}) is hermitean, i.e.\ $C^\dagger = C$, the eigenvalues $\lambda^{(n)}$ are real and we assume them to be positive and non-degenerate, i.e.\ $\lambda^{(0)} > \lambda^{(1)} > \ldots > \lambda^{(N-1)} > 0$.

We can rewrite $\mathbf{v}^{(m) \dagger} C(t) \mathbf{v}^{(n)}$ in two ways by using Eq.\ (\ref{EQN670}) and its hermitean conjugate,
\begin{eqnarray}
 & & \hspace{-0.7cm} \mathbf{v}^{(m) \dagger} C(t) \mathbf{v}^{(n)} = \mathbf{v}^{(m) \dagger} C(t-a) \mathbf{v}^{(n)} \lambda^{(n)} \\
 & & \hspace{-0.7cm} \mathbf{v}^{(m) \dagger} C(t) \mathbf{v}^{(n)} = \mathbf{v}^{(m) \dagger} C(t-a) \mathbf{v}^{(n)} \lambda^{(m)} .
\end{eqnarray}
Since $\lambda^{(m)} \neq \lambda^{(n)}$ for $m \neq n$, we conclude $\mathbf{v}^{(m) \dagger} C(t-a) \mathbf{v}^{(n)} = 0$ for $m \neq n$. In the same way one can show $\mathbf{v}^{(m) \dagger} C(t) \mathbf{v}^{(n)} = 0$ for $m \neq n$. Moreover, it is convenient to normalize the eigenvectors according to \\ $\mathbf{v}^{(n) \dagger} C(t-a) \mathbf{v}^{(n)} = 1$.

Now we consider an arbitrary trial state $| \Psi \rangle$ from the $N$-dimensional space spanned by the states $| \Phi_j \rangle$ included in the correlation matrix $C$,
\begin{eqnarray}
\label{EQN642} | \Psi \rangle = \sum_j a_j | \Phi_j \rangle .
\end{eqnarray}
The corresponding correlation function is
\begin{eqnarray}
C_\Psi(t) = \mathbf{a}^\dagger C(t) \mathbf{a} .
\end{eqnarray}
Since $\mathbf{a}$ (as well as any other complex $N$-component vector) can be expanded in terms of the eigenvectors according to
\begin{eqnarray}
\mathbf{a} = \sum_n \mu^{(n)} \mathbf{v}^{(n)}
\end{eqnarray}
with $\mu^{(n)} \in \mathbb{C}$, we can write for temporal separation $t$
\begin{eqnarray}
\nonumber & & \hspace{-0.7cm} C_\Psi(t) = \sum_{m,n} \mu^{(m) \ast} \mathbf{v}^{(m) \dagger} C(t) \mathbf{v}^{(n)} \mu^{(n)} = \sum_n |\mu^{(n)}|^2 \mathbf{v}^{(n) \dagger} C(t) \mathbf{v}^{(n)} = \sum_n |\mu^{(n)}|^2 \lambda^{(n)} \mathbf{v}^{(n) \dagger} C(t-a) \mathbf{v}^{(n)} = \\
 & & = \sum_n |\mu^{(n)}|^2 \lambda^{(n)}
\end{eqnarray}
and analogously for temporal separation $t-a$
\begin{eqnarray}
C_\Psi(t-a) = \sum_n |\mu^{(n)}|^2 \mathbf{v}^{(n) \dagger} C(t-a) \mathbf{v}^{(n)} = \sum_n |\mu^{(n)}|^2 .
\end{eqnarray}

Now we consider the effective energy corresponding to the correlation function $C_\Psi$,
\begin{eqnarray}
\label{EQN233} E_{\vec{\mu}}^\textrm{eff}(t) = -\frac{1}{a} \log\bigg(\frac{C_\Psi(t)}{C_\Psi(t-a)}\bigg) = -\frac{1}{a} \log\bigg(\sum_m \frac{|\mu^{(m)}|^2}{\sum_n |\mu^{(n)}|^2} \lambda^{(n)}\bigg) .
\end{eqnarray}
Since $0 \leq |\mu^{(m)}|^2 / \sum_n |\mu^{(n)}|^2 \leq 1$ and $\sum_m (|\mu^{(m)}|^2 / \sum_n |\mu^{(n)}|^2) = 1$, the argument of the logarithm in Eq.\ (\ref{EQN233}) is a weighted sum of the eigenvalues, i.e.\ can assume values between the maximal eigenvalue $\lambda^{(0)}$ and the minimal eigenvalue $\lambda^{(N-1)}$. Minimizing $E_{\vec{\mu}}^\textrm{eff}(t)$ with respect to $\vec{\mu}$ is equivalent to maximizing the argument of the logarithm, which corresponds to arbitrary $\mu^{(0)}$ and $\mu^{(1)} = \mu^{(2)} = \ldots = \mu^{(N-1)} = 0$. Thus, $E_{\vec{\mu}}^\textrm{eff}(t)$ is minimized for $\mathbf{a} \propto \mathbf{v}^{(0)}$, which implies $| \Psi \rangle \propto \sum_j v_j^{(0)} | \Phi_j \rangle$.

This is, what we wanted to show: The coefficients of the trial state (\ref{EQN642}) with minimal effective energy at temporal separation $t$ are identical to the components of the eigenvector $\mathbf{v}^{(0)}$ from the GEVP with $t_0 = t-a$.

%%%%%%%%%%%%%%%%%%%%%%%%%%%%%%%%%%%%%%%%%%%%%%%%%%%%%%%%%%%%%%%%%%%%%%%
%%
%%                                   SSS                             
%%                                  S   S                            
%%                                   S                               
%%                                    S                              
%%                                     S                             
%%                                  S   S                            
%%                                   SSS                             
%%
%%%%%%%%%%%%%%%%%%%%%%%%%%%%%%%%%%%%%%%%%%%%%%%%%%%%%%%%%%%%%%%%%%%%%%%

% \newpage

\begin{acknowledgements}

We acknowledge useful discussions with A.\ Ali, J.\ K\"amper, M.\ Pflaumer and G.\ Schierholz.

M.W.\ acknowledges support by the Heisenberg Programme of the Deutsche Forschungsgemeinschaft (DFG, German Research Foundation) -- project number 399217702.

Calculations on the GOETHE-HLR and on the on the FUCHS-CSC high-performance computers of the Frankfurt University were conducted for this research. We would like to thank HPC-Hessen, funded by the State Ministry of Higher Education, Research and the Arts, for programming advice.

\end{acknowledgements}

%%%%%%%%%%%%%%%%%%%%%%%%%%%%%%%%%%%%%%%%%%%%%%%%%%%%%%%%%%%%%%%%%%%%%%%
%%
%%                                   SSS                             
%%                                  S   S                            
%%                                   S                               
%%                                    S                              
%%                                     S                             
%%                                  S   S                            
%%                                   SSS                             
%%
%%%%%%%%%%%%%%%%%%%%%%%%%%%%%%%%%%%%%%%%%%%%%%%%%%%%%%%%%%%%%%%%%%%%%%%

\bibliographystyle{apsrev4-1}
\bibliography{literature.bib}

\end{document}